\documentclass[prl,twocolumn,showpacs,superscriptaddress]{revtex4}

\usepackage{graphicx}
\usepackage{dcolumn}
\usepackage{bm}
\usepackage{epsfig}
\usepackage{footnote}
\usepackage{times}

\begin{document}

\title{Long-Distance Decoy-State Quantum Key Distribution in Optical Fiber}

\author{Danna Rosenberg}
\affiliation{Applied Modern Physics, MS D454, Los Alamos National
Laboratory, Los Alamos, NM 87545}
\author{Jim W. Harrington}
\affiliation{Applied Modern Physics, MS D454, Los Alamos National
Laboratory, Los Alamos, NM 87545}
\author{Patrick R. Rice}
\affiliation{Applied Modern Physics, MS D454, Los Alamos National
Laboratory, Los Alamos, NM 87545}
\author{Philip A. Hiskett}
\affiliation{Applied Modern Physics, MS D454, Los Alamos National
Laboratory, Los Alamos, NM 87545}
\author{Charles G. Peterson}
\affiliation{Applied Modern Physics, MS D454, Los Alamos National
Laboratory, Los Alamos, NM 87545}
\author{Richard J. Hughes}
\affiliation{Applied Modern Physics, MS D454, Los Alamos National
Laboratory, Los Alamos, NM 87545}
\author{Adriana E. Lita}
\affiliation{National Institute of Standards and Technology, 325
Broadway, Boulder, CO 80305}
\author{Sae Woo Nam}
\affiliation{National Institute of Standards and Technology, 325 Broadway, Boulder, CO 80305}
\author{Jane E. Nordholt}
\affiliation{Applied Modern Physics, MS D454, Los Alamos National
Laboratory, Los Alamos, NM 87545}
\date{Submitted 26 July 2006}

\begin{abstract}
The theoretical existence of photon-number-splitting attacks creates
a security loophole for most quantum key distribution (QKD)
demonstrations that use a highly attenuated laser source.  Using
ultra-low-noise, high-efficiency transition-edge sensor
photo-detectors, we have implemented the first version of a decoy
state protocol that incorporates finite statistics without the use
of Gaussian approximations in a one-way QKD system, enabling the
creation of secure keys immune to photon-number-splitting attacks
and highly resistant to Trojan horse attacks over $107$~km of
optical fiber.
\end{abstract}


\maketitle

Quantum key distribution (QKD), which enables users to create a
shared key with secrecy guaranteed by the laws of physics
\cite{roadmap}, is arguably the most advanced application in the
growing field of quantum information science. Since the first
demonstration in $1992$ \cite{BBBSS92}, the field has advanced
sufficiently that commercial systems are now available. Most current
QKD implementations use ``prepare and measure'' protocols that
involve the sender (Alice) preparing a single photon in a quantum
state and sending it to the receiver (Bob), who then measures the
photon. Attempts by an eavesdropper (Eve) to obtain information
about the state of the single photon will introduce an error rate in
the transmission, which alerts the users to Eve's presence.

For example, to implement the Bennett-Brassard $1984$ (BB84)
protocol \cite{BB84}, Alice randomly encodes a single photon with
either a $0$ or a $1$ in one of two conjugate bases and sends the
photon to Bob. Bob performs a measurement in one of the two bases,
and communicates the time slots for which he obtained detection
events. Alice and Bob then create a sifted key by only retaining
events where they used the same basis. Ideally, Alice's sifted bits
should be perfectly correlated with Bob's if Eve did not attack the
transmission, but any real system has error rates due to
experimental imperfections. Error correction \cite{Brassard94}
removes these errors, leaving Alice and Bob with a perfectly
correlated key. However, this key is not yet completely secret
because, in principle, the errors may have arisen from Eve attacking
the system. Therefore, a final step of privacy amplification
\cite{Bennett95} is used to obtain a shorter, secret key about which
Eve has negligible information.

The lack of readily available single-photon sources, especially at
telecom wavelengths where most fiber-based QKD systems operate,
modifies the simple picture outlined above considerably. If the
source emits more than one photon, Eve could remove one of the
photons and store it until Bob announces his basis choice, at which
time she would measure the photon in the correct basis and learn the
bit value without introducing any errors. Therefore, in addition to
assuming that all errors arise from Eve's interaction with single
photons, it is also necessary to assume that Eve can gain full
information about any sifted bits that arose from multi-photon
events. To determine the number of sifted bits that were encoded in
single photons, it is often assumed that the transmission channel
acts as a simple beamsplitter \cite{BBBSS92}. However, an
eavesdropper with unlimited technological capabilities may modify
the channel properties so that this is no longer valid. For
instance, she may perform a photon-number-splitting (PNS) attack by
replacing the link with a lossless channel, blocking as many single
photons at the output of Alice that she can, while keeping the rate
of photons that Bob receives constant, and removing one photon from
each multiphoton pulse \cite{Brassard00}. Protection against such
attacks requires far more privacy amplification than the case where
a beamsplitter channel is assumed, and if the rate of multiphotons
present at the output of Alice is greater than the rate of detection
events recorded by Bob, then Eve could have full knowledge of every
sifted bit.

QKD systems often use heavily attenuated laser sources, which
results in a Poisson distribution of photon number. The fraction of
non-vacuum pulses that contain more than one photon is approximately
$\mu/2$ when the laser is pulsed with a mean photon number $\mu<1$.
To keep the rate of multiphotons sufficiently low for PNS security,
it is necessary to operate with $\mu$ on the order of the channel
transmittance $\eta$, yielding a sifted bit rate that is
proportional to $\eta^2$ \cite{Lutkenhaus00}.  As the transmission
loss increases and the sifted bit rate decreases, detector dark
counts play an increasingly important role, eventually leading to
such high error rates that secret key generation is impossible.  For
fiber QKD, where the channel transmittance drops off exponentially
with distance, the requirement of PNS security was until recently
thought to severely limit the link length for weak coherent pulse
QKD \cite{Shields_pns04,Hiskett_200km}.

The recent development of decoy state protocols
\cite{Hwang_decoy,Lo05,Wang05,Harrington05} has drastically improved
the outlook for the security of weak laser based QKD.  Decoy-state
QKD allows the users to place a rigorous lower bound on the single
photon channel transmittance, including receiver losses, and
therefore the number of detections at Bob that originated from
single photons. Because no assumptions are made about modifications
to the channel transmittance by an eavesdropper, a PNS attack would
easily be detected.  Decoy-state QKD has previously been
demonstrated over link lengths of $15$ and $60$~km
\cite{Lo_15km,Lo_60km}, with a suggested maximum PNS-secure range of
about $140$~km if InGaAs avalanche photodiodes with the
best-reported parameters for QKD in the literature are used
\cite{lo_pra05}. However, those experiments employed a two-way
system that has been shown to be susceptible to Trojan horse attacks
\cite{Gisin02}, negating the purpose of QKD to create
unconditionally secure keys. In contrast, the present work was
performed with a one-way system which is much less susceptible to
Trojan horse attacks \footnote {It has recently been shown that, due
to finite backscattering of experimental components, an eavesdropper
can obtain some information by probing a one-way system
\cite{Gisin06}, but the technological capabilities needed to perform
such an attack are far greater than would be necessary for a two-way
system.}. In this paper, we report on the first experimental
decoy-state QKD demonstration in a one-way QKD system that can
create unconditionally secure quantum key.

The simplest decoy state protocol requires Alice to emit signals
whose $\mu$ values are randomly toggled between two values $\mu_1$
and $\mu_0$.  For a given signal, Eve does not know whether Alice
used $\mu_0$ or $\mu_1$, so she must treat single photon signals
from either mean photon number identically. Because the fraction of
single photon signals depends on $\mu$, it is impossible for Eve to
perform a PNS attack by simultaneously modifying the channel
transmission correctly for more than one value of $\mu$. By
comparing the number of detection events from $\mu_0$ and $\mu_1$
transmissions, Alice and Bob are able to place strict bounds on the
single photon transmittance of the channel.

A three-level decoy-state protocol ([$\mu_0, \mu_1, \mu_2=0$]) with
$\mu_1\ll\mu_0$ enables even better characterization of the channel
parameters, which can be illustrated as follow.  Bob's count of
detection events when Alice sent vacuum ($\mu_2$) provides an
estimate of the background and dark count detection probability,
$y_0$, per clock cycle of the system. From this estimate, they can
develop upper and lower bounds on $y_0$ with a user-defined level of
confidence $1 - \epsilon$, with $\epsilon\ll1$. The confidence
interval calculations in our case were computed numerically as
opposed to making a Gaussian approximation, which may be a poor fit
far out in the tails of the binomial distribution governing both the
transmission and error probabilities. Next, they consider how many
detection events Bob received when Alice prepared mean photon number
$\mu_1 \ll 1$. After subtracting off background, most of the
remaining events are from single-photon signals, providing an
estimate and confidence levels for the single photon transmittance
$y_1$.  Finally, they can utilize the lower bound on $y_1$ to
determine the number of the stronger $\mu_0$ detection events that
originated as single photon signals at Alice. While this outline is
helpful for gaining intuition, it does not explain the specific
values of mean photon numbers that should be chosen for an
experiment such as ours.

More generally, the channel analysis is carried out by
simultaneously solving for the $n$-photon signal transmittance
variables $y_n$ under a set of linear inequalities formed by
confidence intervals $[Y_j^-,Y_j^+]$ on the detection probabilities
per clock cycle $Y_j$ for each $\mu_j$ \cite{Harrington05}:
$Y_j^- ~\leq~ e^{-\mu_j} \sum_{n=0}^{\infty}
\frac{\left(\mu_j\right)^n}{n!}y_n ~\leq~ Y_j^+$.
The region of consistent solutions forms a convex polyhedron, and a
lower bound $y_1^-\leq y_1$ can easily be found by linear
programming. A similar set of inequalities relate the confidence
intervals on the observed bit error rates for each $\mu_j$ to the
$y_n$ and the $n$-photon bit error rates $b_n$. While simultaneously
solving both sets of inequalities could in principle yield a tight
bound on the single photon bit error rate $b_1$, we chose to instead
use a conservative upper bound $b_1^{+}$ by treating all observed
sifted bit errors as having come from single-photon signals. Details
on channel estimation and optimization of experimental parameters
for fiber decoy state QKD will be published separately
\cite{Harrington_tbp}.

\begin{figure}[htb]
\includegraphics[width=3.375in]{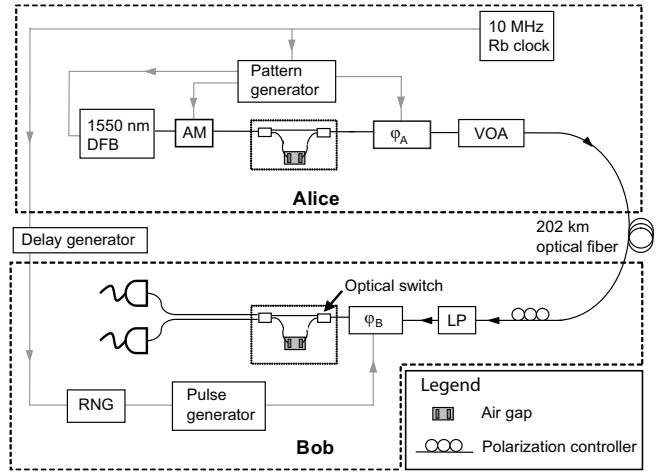}
\caption{QKD system used in this work.  DFB: distributed feedback
laser; VOA: variable optical attenuator; AM: amplitude modulator;
LP: linear polarizer; RNG: random number generator.}
 \label{fig:system}
\end{figure}

The switched interferometer QKD system used in this work was
identical to that described in detail elsewhere
\cite{Hiskett_systems,Hiskett_200km}, except for the addition of an
amplitude modulator in Alice, which was used to produce the
different decoy state signal strengths.  As shown in Fig.
\ref{fig:system}, the system was composed of a phase encoding
switched interferometer and low-noise, high efficiency single-photon
sensitive superconducting transition-edge sensors (TESs)
\cite{Irwin95,PRA05}. Synchronization for both Alice and Bob was
achieved through the use of a single clock, making the system
impractical for use outside the laboratory, but straightforward
modifications will yield a system with separate clocks using quantum
clock recovery techniques \cite{Hughes05}.  A pattern generator
pre-loaded with a random bit file provided bit and basis selection
for Alice, but in a practical system cryptographically strong random
number generators would provide the selection \cite{Hughes05}. These
two relatively minor modifications to the system will be implemented
in the near future. In contrast to our previous work using TESs in a
phase encoded system \cite{APL06,Hiskett_systems,Hiskett_200km}, in
which we used one detector and time-multiplexed the signals at Bob's
phase decoder, here we used two detectors to enable operation at a
higher clock rate (2.5 MHz for this experiment). The detectors had
fiber-coupled system efficiencies of $33$\% and $50$\%, which were
lowered from the detector value of $89$\% by the inclusion of
filters to reduce the rate of blackbody radiation reaching the
detectors. This imbalance between the two detectors reduces the
entropy of the raw key, which must be accounted for during privacy
amplification. The background rate of detection events, set by
blackbody radiation, was $3$ counts per second. The timing jitter of
the detectors was $100$~ns FWHM, and the thermal recovery time was
$4~\mu$s.  The system transmitted over a $202$~km link of dark
optical fiber, and shorter distances were obtained by redefining
Alice's enclave to include some length of the optical fiber
\cite{Hiskett_200km}.  Redefining the system in this way simply
means that Alice has an extra attenuator composed of fiber that
lowers the mean photon number exiting her enclave. Therefore, our
mapping to shorter distances is completely equivalent to using a
shorter length of fiber.

We implemented a decoy-state BB84 protocol using three levels of
$\mu$: a high $\mu_0$, a moderate $\mu_1$, and a low $\mu_2$ that
approximates the vacuum state.  The probabilities of sending
$\mu_0$, $\mu_1$, or $\mu_2$ were $83.1$\%, $12.3$\% and $4.6$\%,
respectively.  Near-optimal $\mu$ values and probabilities were
obtained by performing simulations to maximize the secret bit rate
for various channel parameters. Because of the finite extinction
ratio of the amplitude modulator, $\mu_2$ was not zero but was
instead less than $1.0$\% of $\mu_0$. Use of a small nonzero value
for $\mu_2$ results in slightly worse bounds on the single photon
transmission, and this effect was included in our analysis. The
user-defined confidence parameter for each bound was chosen to be
$\epsilon = 10^{-7}$, resulting in a final key of which, with
probability greater than $1-6\times10^{-7}$, Eve knows less than one
bit.

After sufficient data were collected, the bits arising from pulses
at $\mu_0$ were sifted, error corrected, and privacy amplified.
After sifting, the bits were shuffled to permute the errors and make
error correction more efficient. In addition, half of the bits,
randomly chosen, were flipped by both Alice and Bob to ensure that
the final key had an equal distribution of zeros and ones. Error
correction was performed using the modified CASCADE algorithm
\cite{modCASCADE}, which has an efficiency of 7--13\% over the
Shannon limit.  We performed privacy amplification using Toeplitz
matrix universal hash functions \cite{Toeplitz} to provide
protection against arbitrary basis-independent attacks \cite{GLLP},
yielding a total of $N_{\rm sec}$ secret bits:
\begin{eqnarray*}
N_{\rm sec} &=& s \left\lbrack 1 - H_2 \left( b_1^+ \right)
\right\rbrack - N_{\rm sift} \left\lbrack f_{\rm ec} H_2 \left( B
\right) +  \left( 1-H_2 \left( z \right) \right) \right\rbrack
\end{eqnarray*}
where $N_{\rm sift}$ is the number of sifted bits, $s$ is the
calculated lower bound on the number of single photons present in
the sifted key, $b_1^+$ is the calculated upper bound on the single
photon error rate, $f_{\rm ec}$ is the efficiency of the error
correction protocol relative to the Shannon limit, $B$ is the
observed error rate for all signals that enter the sifted key, $z$
is the fraction of zeros in the sifted key before half the bits were
flipped, and $H_2$ is Shannon entropy.

\begin{figure}[tb]
\includegraphics[width=2.75in]{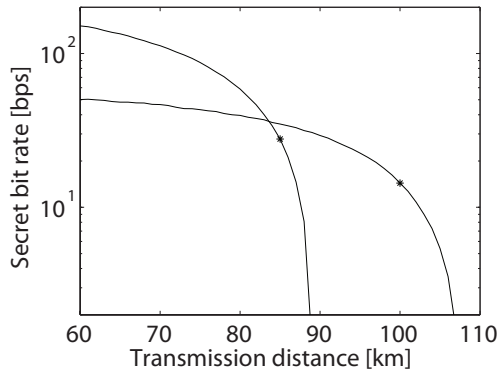}
\caption{Secret bit rate (bps denotes bits per second) vs
transmission distance for the two experimental data sets. The
asterisks mark the transmission distances quoted in the text. Longer
distances could be achieved in this system if different mean photon
numbers are used.}
 \label{fig:rsec_vs_dist}
\end{figure}

We collected data at two different sets of values of $\mu$, one
selected for transmission at $85$~km (corresponding to $117$~km of
fiber being defined as residing within Alice) and the other for
$100$~km (corresponding to $102$~km of fiber residing within Alice's
enclave). For each data set, timing windows for accepting detected
events were chosen to maximize the secret bit rate
\cite{APL06,Hiskett_systems}. From the first data set, using mean
photon numbers at the exit of Alice's enclave of [$\mu_0$, $\mu_1$,
$\mu_2$] = [$0.487$, $0.0639$, $1.05\times10^{-3}$] at $85$~km, we
created $9.9\times 10^3$ secret bits in 351 s from $2.2\times 10^5$
sifted bits using $120$~ns windows. From the second data set, which
used mean photon numbers [$0.297$, $0.099$, $2.75\times10^{-3}$] at
$100$~km, we generated a $1.2\times 10^4$ secret bits from
$1.9\times 10^5$ sifted bits collected over $828$~s with $220$~ns
windows.  The observed error rates at $\mu_0$ for the two data sets
were $3.3$\% and $4$\%, consistent with the expected error rate due
to interferometer visibility and background counts.  The lower
bounds on the fraction of sifted bits that originated as single
photons were $0.46$ and $0.55$, compared to $0.61$ and $0.74$ for a
beamsplitter channel. The number of secret bits generated is less
than the number of non-PNS-secure bits that would have been
generated at $\mu_0$ by assuming a random deletion channel
($4.4\times10^4$ and $4.9\times10^4$ at $85$~km and $100$~km,
respectively), but those numbers assume that Eve is unable to modify
the channel properties. Consequently, the secret bits generated
using our decoy state protocol are immune to PNS attacks, whereas
they would not be PNS secure under the beamsplitter channel
assumption.

Even though the $\mu$ values were chosen to be near-optimal for
particular link lengths, we can analyze the results over other
distances by redefining the system so that Alice's enclave includes
a different amount of the $202$~km optical fiber link
\cite{Hiskett_200km}. Figure \ref{fig:rsec_vs_dist} shows the secret
bit rate as a function of transmission distance. For the data set
optimized for $100$~km, we find that a secret key can be exchanged
over 107 km of optical fiber.  Considerably longer ranges of
$150$--$200$~km should be possible in this system by using different
$\mu$ values optimized for longer distances.

\begin{figure}[htb]
\includegraphics[width=3.375in]{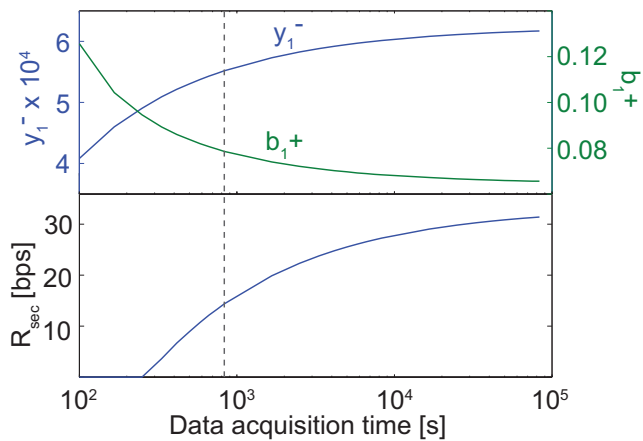}
\caption{Single photon transmittance $y_1^-$, error rate $b_1^+$,
and secret bit rate dependence on data acquisition time at $100$~km.
The dashed line indicates the actual data acquisition time of 828 s
at which secret bits were generated; the other points on the graph
are estimates based on the data.}\label{fig:rsec_vs_time}
\end{figure}

Because the extent to which the single photon transmittance can be
bounded is dependent on the photo-count statistics, acquiring data
for longer times will result in not only more secret bits, but also
a higher {\it rate} of secret bit production.  Figure
\ref{fig:rsec_vs_time} displays the results of a simulation of
longer acquisition times.  In general, the bound on single photon
transmittance does not depend on whether the quantum channel is
stationary, but for the simulation we assume that Eve does not vary
her attack. For a given confidence parameter, longer acquisition
times result in a tighter lower bound on the single photon
transmittance and a tighter upper bound on the single photon sifted
bit error rate, leading to a higher secret bit rate. For this
simulation, we have not adjusted the mean photon numbers used when
the data acquisition time is increased; re-optimization of the mean
photon numbers for longer times is expected to increase the secret
bit rate even further.

By incorporating low-noise transition-edge sensors into a one-way
QKD system and implementing a three-level decoy protocol, we were
able to generate key secure against PNS attacks and with only very
limited susceptibility to Trojan horse attacks over $107$~km of
optical fiber. This distance far surpasses the previous maximum
PNS-secure transmission distance of $67.5$~km that used very weak
mean photon numbers rather than the decoy state protocol in
essentially the same system \cite{Hiskett_200km}. In contrast to
other work, this demonstration was the first to implement a
finite-statistics protocol to bound the channel transmittances
without resorting to Gaussian approximations. We used a conservative
method to estimate the error rate on single photon signals, but
future work may incorporate tighter bounds on the single photon
error rate, resulting in higher secret bit rates and longer ranges.
System clock rates as much as five times higher are expected to be
achieved with improvements in the detector readout electronics,
leading to higher secret bit rates. Based on the results of
simulations, we expect that this system is capable of PNS-secure
decoy-state QKD over $150$--$200$~km of optical fiber, and
improvements in filtering of the ambient blackbody photons could
increase this distance even further to $250$~km or more.

{\it Note added:} Recently, we became aware of similar work
performed elsewhere \cite{chinadecoy06}.

The authors thank DTO and the NIST Quantum Initiative for financial
support, Alan Migdall for the loan of an optical switch, and Joe
Dempsey and Corning Inc. for loaning the $202$~km of single-mode
optical fiber.

\end{document}